\documentclass{article}
\usepackage[utf8]{inputenc}

\usepackage{geometry}
\usepackage{amsmath}
\usepackage{latexsym}
\usepackage{amsfonts}
\usepackage{mathrsfs}
\usepackage{upgreek}
\usepackage{graphics}
\usepackage{graphicx}
\usepackage{epstopdf}
\usepackage{epsfig}
\usepackage{amsbsy}
\usepackage{amsfonts}
\usepackage{amsmath}
\usepackage{amssymb}
\usepackage{bm}
\usepackage{color}
\usepackage{setspace}
\usepackage{authblk}

\title{Morphology of multiple constant height hydraulic fractures versus propagation regime}

\author{E.V. Dontsov$^1$}
\date{%
    $^1$egor@resfrac.com, ResFrac Corporation, Palo Alto, USA
}

\begin{document}

\maketitle
\begin{abstract}
    \noindent The purpose of this study is to investigate morphology of simultaneously propagating hydraulic fractures. Simultaneous propagation of hydraulic fractures occurs during stimulation of horizontal wells, and, in particular, several initiation points or perforation intervals along the well are often used to promote the growth of multiple hydraulic fractures at the same time. Numerical simulations demonstrate that there are situations, in which stress interaction between the fractures is minimal and this results in a very similar geometry of each fracture. At the same time, for some other parameters, the stress interaction is very strong, so that the fractures interact with each other and develop complex shapes. By focusing on the constant height hydraulic fractures, it is shown that there is a dimensionless parameter that controls such a behavior. In particular, fractures propagating in the toughness dominated regime lead to complex shapes, while when fluid viscosity dominates the response, then the fractures are more regular and uniform. A series of numerical examples is presented to illustrate the findings.
\end{abstract}

\section{Introduction}

Multistage hydraulic fracturing is often used to enhance recovery of hydrocarbons from low permeability formations~\cite{King2012}. To optimize the process, multiple hydraulic fractures are initiated simultaneously from a horizontal wellbore. In addition, this is usually supplemented by a limited entry design, which promotes uniform flow distribution between the fractures. It was shown, however, that even in the simplest case of no geological layers, such a problem can lead to the development of instabilities that can cause complex fracture shapes~\cite{Dont2020a}. In addition, similar behavior was observed for field scale simulations of hydraulic fracturing~\cite{McClure2020}. This behavior is important from both fundamental and practical points of view. First, it allows to better understand processes happening underground, and second, it permits to control this behavior using operational procedures.

As was shown in~\cite{Dont2020a}, hydraulic fracture propagation regime is the key factor that determines morphology of simultaneously propagating hydraulic fractures. For simple fracture geometries, such as plane strain, radial, and constant height cracks, there are four limiting cases or propagation regimes~\cite{Detou2016}. These are determined by the competition between viscous dissipation and fracture toughness, as well as the fluid storage inside the fracture or the surrounding formation. It is important to note that these mathematical conclusions are based on the model that features linear elastic rock deformation, Newtonian fracturing fluid, linear elastic fracture mechanics, Carter's leak-off, and uniform injection rate. Other important physical processes, such as proppant transport or the effect of hydrostatic pressure are ignored. Nevertheless, it was concluded in~\cite{Dont2020a} that when viscosity dominates, then all the fractures are nearly identical. At the same time, when toughness dominates, then the fractures strongly compete with each other and develop petal-like shapes. This was the case for ``radial'' geometry, or when there were no geological layers. Qualitatively similar observations were made for field scale hydraulic fracture modeling in~\cite{McClure2020}. Namely, the increase of fracture toughness leads to more fracture asymmetry, while the increase of viscosity (and/or decrease of toughness) leads to more regular and symmetric fractures. One of the challenges, however, is to quantitatively determine the parameter that governs such a transition in fracture morphology. As a result, this study aims to extend the previous analysis~\cite{Dont2020a} for the case of constant height hydraulic fractures. While such a geometric assumption is still far from a general field scale simulation, it is arguably a better approximation relative to the radial case, especially when strong barriers limit fracture propagation in the vertical direction. Finally, and most importantly, it is possible to perform a relatively simple analysis for the constant height hydraulic fractures, which is an essential component that is needed to quantify the fracture morphology.

Constant height hydraulic fractures are often referred to as Perkins–Kern–Nordgren (PKN)~\cite{Perk1961,Nord1972} fractures. Many researchers have investigated propagation of PKN fractures from various perspectives. A few papers that are relevant to this work are~\cite{Nolte1991,Sarva2015, Dont2016}. The classical PKN model does not account for the effect of fracture toughness and therefore is unsuitable for analyzing the problem under consideration since the transition from toughness domination to viscosity domination presents the most interest. The first attempt to include the effect of toughness was done in~\cite{Nolte1991}, where the propagation condition in terms of pressure was taken from the uniformly pressurized radial fracture with the diameter equal to the reservoir height. This methodology was then revised in~\cite{Sarva2015}, in which the authors proposed a more accurate procedure that utilizes energy considerations. Finally, it was shown in~\cite{Dont2016} that the second energetic approach is indeed more accurate (albeit quite marginally) by comparing the results to the solution with non-local elasticity as well as to the simulations performed with a fully planar model.

One prerequisite for conducting the study on simultaneous propagation of multiple hydraulic fractures is the knowledge of the parametric space for the problem. For instance, for the case of radial geometry~\cite{Dont2020a}, the prerequisite paper is~\cite{Dont2016f}, which builds on top of many previous publications and outlines boundaries in the parametric space that quantify dominance of various physical processes. In the context of constant height or PKN fractures, such a parametric space was recently constructed in~\cite{Dont2022}. In particular, the model that accounts for toughness based on the results in~\cite{Sarva2015} is used due to its simplicity and still admissible accuracy. Section~\ref{sec2} summarizes mathematical formulation for the model and describes the aforementioned parametric space. Equipped with this parametric space, several cases are constructed for multiple fractures that lie within different regions of the parametric space. Numerical simulations are performed using commercial code ResFrac\footnote{www.resfrac.com}~\cite{McClure2018} and results of simulations for simultaneous propagation of multiple hydraulic fractures are presented in section~\ref{sec3} for different regimes. Finally, conclusions are summarized in section~\ref{secsumm}.

\section{Parametric analysis of a single constant height hydraulic fracture}\label{sec2}

\begin{figure}[h]
\centering\includegraphics[width=0.65\linewidth]{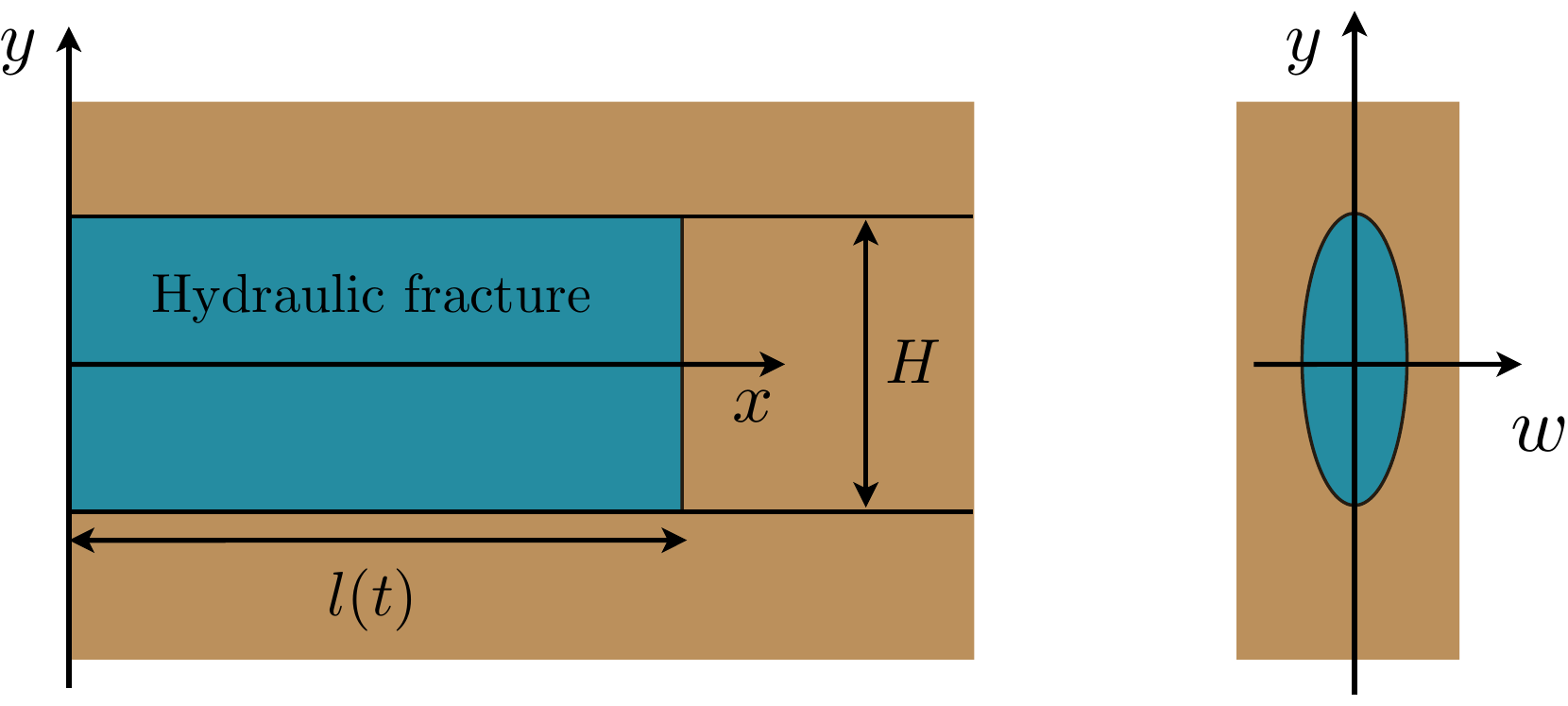}
\caption{Schematics of one wing of a constant height hydraulic fracture.}
\label{figschem}
\end{figure}

This section outlines the mathematical model for a single PKN fracture and also presents the parametric space for the problem. With the reference to Fig.~\ref{figschem}, let the fracture be vertical and occupy the $(x,y)$ plane, where $y$ is the vertical coordinate and $x$ is the horizontal coordinate. The fracture height is denoted by $H$, while the half-length is $l(t)$. The model assumes that $H\ll l$, which ensures that the flow is predominantly horizontal and also confirms validity of the local elasticity assumption.

By following the classical PKN model~\cite{Perk1961,Nord1972}, fracture width in each vertical fracture cross-section is assumed to be elliptical, and the fracturing fluid pressure is determined based on the local plane strain elasticity assumption:
\begin{equation}\label{PKNrel}
w(x,y)~=~\dfrac{4}{\pi} \bar w(x) \sqrt{1-\Bigl(\dfrac{2y}{H}\Bigr)^2},\qquad p(x)~=~\dfrac{2 E' \bar w(x)}{\pi H},\qquad \bar w(x)=\dfrac{1}{H}\int_{-H/2}^{H/2} w(x,y)\,dy.
\end{equation}
Here $w(x,y)$ is the fracture opening, $\bar w(x)$ is the effective or averaged width, $p(x)$ is the fluid pressure, and $E'=E/(1\!-\!\nu^2)$ is the plane strain Young's modulus. The primary governing equation is the vertically-averaged volume balance, which reads:
\begin{equation}\label{lubrication}
\dfrac{\partial \bar w}{\partial t}+\dfrac{\partial \bar{q}_x }{\partial x} + \dfrac{C'}{\sqrt{t\!-\!t_0(x)}}~=~\dfrac{Q_0}{H} \delta(x),\qquad \bar{q}_x~=~-\frac{1}{12H\mu} \dfrac{\partial p}{\partial x} \int_{-H/2}^{H/2}w^3\, dy~=~-\frac{\bar w^3}{\pi^2\mu} \dfrac{\partial p}{\partial x}.
\end{equation}
In the above equation $\mu$ is the fluid viscosity, $\bar q_x$ denotes the horizontal flux, $C'=2C_l$ is the scaled Carter's leak-off coefficient, and $t_0(x)$ corresponds to time when the fracture front was located at the point $x$.

Propagation condition at the lateral fracture tip is based on the results from~\cite{Sarva2015} and can be written in terms of the toughness dependent pressure condition as
\begin{equation}\label{PKNpbc}
p(l)~=~\dfrac{2 K_{Ic}}{\sqrt{\pi H}},
\end{equation}
where $K_{Ic}$ is fracture toughness. This propagation condition is supplemented by the zero flux at the tip, which translates into the following equation for fracture length:
\begin{equation}\label{lengtheq}
    \dfrac{dl}{dt} = \dfrac{\bar q_x(l)}{\bar w(l)} = -\dfrac{2 E'}{3\pi^3 \mu H}\dfrac{\partial \bar w^3}{\partial x}\biggr|_{x=l}.
\end{equation}

By following the results in~\cite{Dont2022}, one possibility to reduce the number of parameters in the problem is to introduce the following normalization:
\begin{equation}\label{normalization}
    \Omega = \dfrac{\bar w}{w_*},\qquad \lambda = \dfrac{l}{l_*},\qquad \tau = \dfrac{t}{t_*},\qquad \xi=\dfrac{x}{l}.
\end{equation}
Here the width, length, and time scales are given by
\begin{equation}\label{scales}
    w_* = \dfrac{(\pi H)^{1/2}\, K_{Ic}}{E'},\qquad l_* = \dfrac{H^2 K_{Ic}^4}{2\pi E'^3\mu Q_0} ,\qquad t_* = \dfrac{H^{7/2} K_{Ic}^5}{2 \pi^{1/2} E'^4\mu Q_0^2}.
\end{equation}
By adopting such a normalization, the governing equations~(\ref{PKNrel})--(\ref{lengtheq}) become
\begin{equation}\label{PKNdim}
\dfrac{\partial \Omega}{\partial \tau}-\dfrac{\xi\dot\lambda}{\lambda}\dfrac{\partial \Omega}{\partial \xi} - \dfrac{1}{\lambda^2}\dfrac{\partial ^2 \Omega^4} {\partial \xi^2}+\dfrac{\phi}{\sqrt{\tau\!-\!\tau_0(\xi)}}~=~ \delta(\xi),\qquad \Omega(1) = 1,\qquad \dfrac{d\lambda}{d\tau} = -\dfrac{4}{3\lambda}\dfrac{\partial \Omega^3}{\partial \xi}\biggr|_{\xi=1},
\end{equation}
where the dimensionless parameter that quantifies leak-off is
\begin{equation}\label{phidef}
    \phi = \Bigl(\dfrac{H^{5} K_{Ic}^{6} C'^4}{4 \pi^3 E'^4\mu^2 Q_0^4} \Bigr)^{1/4}.
\end{equation}
Thus, the solution to the problem in terms of length and effective width depends on two dimensionless parameters, namely dimensionless time $\tau$ and dimensionless leak-off $\phi$. Conceptually, these parameters are similar to those for a radial fracture~\cite{Dont2016f}, but quantitatively they are very different.

\begin{figure}[h]
\centering\includegraphics[width=0.45\linewidth]{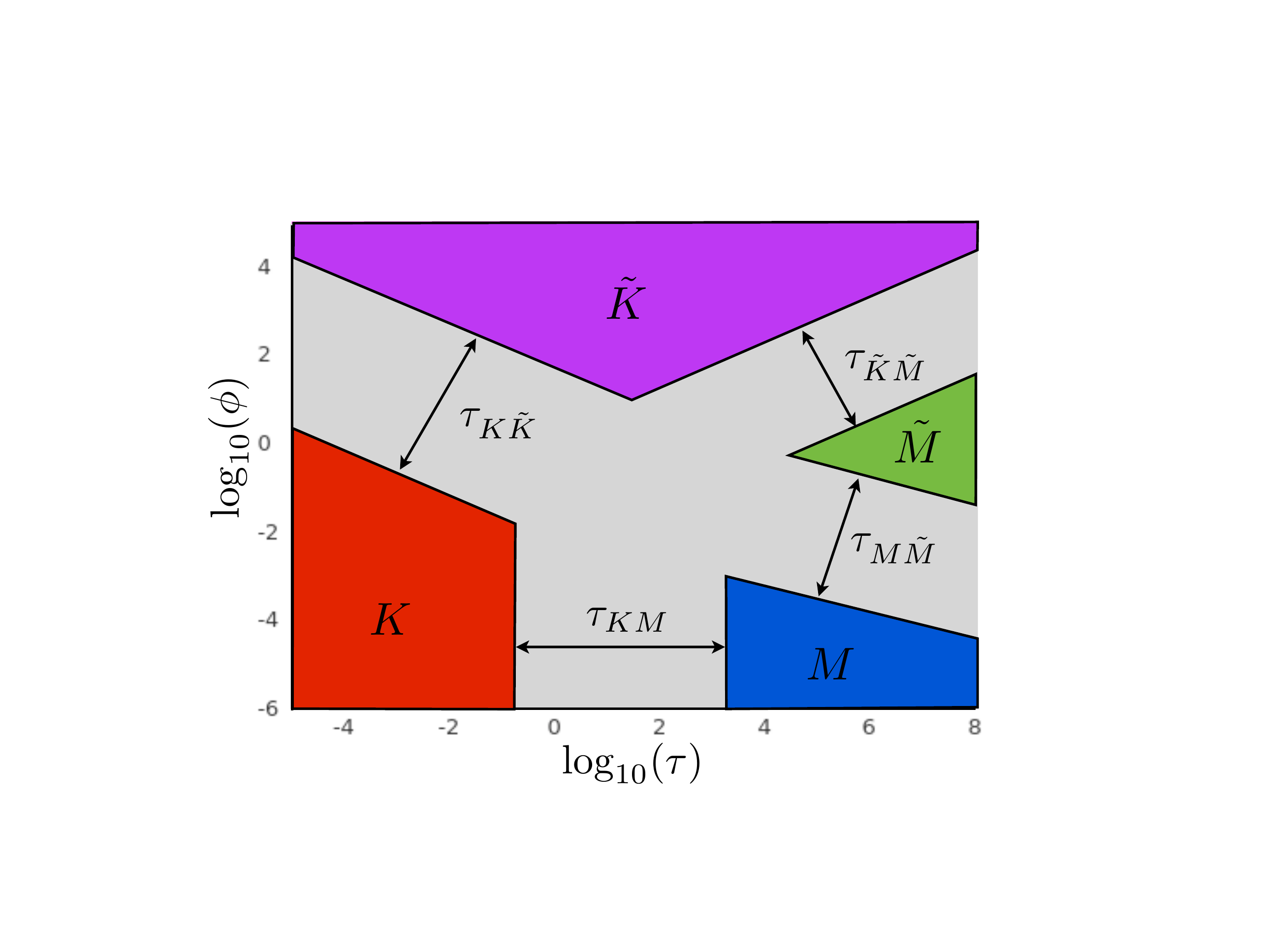}
\caption{Parametric space for a single constant height hydraulic fracture.}
\label{figparam}
\end{figure}

Parametric space for this problem was constructed in~\cite{Dont2022} and is shown in Fig.~\ref{figparam}. The blue, red, green, and magenta regions indicate zones of applicability of storage viscosity or $M$, storage toughness or $K$, leak-off viscosity or $\tilde M$, and leak-off toughness or $\tilde K$ solutions, respectively. These are the zones, in which the difference between the ``true'' solution and the limiting self-similar solution falls below a small threshold, see~\cite{Dont2022} for more details. Here the storage viscosity limit corresponds to dominance of viscosity over toughness and small leak-off, i.e. when most of the injected fluid stays inside the fracture. In contrast, storage toughness limit occurs when toughness dominates over the viscosity, while most of the injected fluid still stays inside the fracture. The leak-off regimes are defined in a similar fashion, but most of the injected fluid leaks into the formation in these limits. Equations that define boundaries of the transition regions are given by
\begin{eqnarray}\label{PKNtransitions}
\tau_{MK} &=& \tau, \qquad\tau_{MK,1} = 0.11,\qquad\tau_{MK,2} = 2.3\times 10^3,\notag\\
\tau_{K\tilde K}&=& \tau \phi^{2},\qquad \tau_{K\tilde K,1} = 5.7\times 10^{-5},\qquad\tau_{K\tilde K,2} = 3.1\times 10^{3},\notag\\
\tau_{\tilde K \tilde M}&=&\tau \phi^{-2},\qquad \tau_{\tilde K\tilde M,1} = 0.18,\qquad\tau_{\tilde K\tilde M,2} = 6.5\times 10^{4},\notag\\
\tau_{M \tilde M} &=& \tau\phi^{10/3},\qquad \tau_{M\tilde M,1} = 2.0\times 10^{-7},\qquad\tau_{M\tilde M,2} = 2.9\times 10^{3}.
\end{eqnarray}
In the context of the analysis of propagation of multiple hydraulic fractures, this parametric space allows to select problem parameters that correspond to dominance of either toughness, viscosity, or fall into the situation when both are important.

\section{Numerical results for multiple fractures}\label{sec3}

A series of numerical simulations for simultaneous propagation of multiple hydraulic fractures is performed in ResFrac, which is a fully coupled 3D hydraulic fracturing and reservoir simulator~\cite{McClure2018}. Three basic cases are considered: toughness domination (denoted by $K$), transition (denoted by $MK$), and viscosity (denoted by $M$) domination. The input parameters as well as the corresponding dimensionless parameters $\tau$ and $\phi$ are summarized in Tab.~\ref{tab1}, see~(\ref{normalization}), (\ref{scales}), and~(\ref{phidef}) for the definitions of $\tau$ and $\phi$. Note that leak-off cases are not considered because previous study~\cite{Dont2020a} showed that leak-off does not qualitatively affect the behavior. Uniform flux distribution is achieved by substantially increasing perforation friction, the effect of gravity is neglected, and the height constraint is introduced by having strong barriers above and below the primary layer of interest. All the injection points are located in the middle of the layer. Only 5 fractures are initiated, all of which are planar and perpendicular to the well in the base case. Also, the base simulations have 10~m uniform spacing between the fractures, while other values are used to investigate the effect of spacing, namely 20, 40, 80, and 160 m.

\begin{table}
\begin{center}
\begin{tabular}{ |c|c|c|c| } 
    \hline
    Property & $K$ & $MK$ & $M$ \\
    \hline
    $K_{Ic}$~[MPa$\cdot$m$^{1/2}$] & 5 & 3 & 3 \\ 
    $H$~[m] & 80 & 80 & 80 \\ 
    $E$~[GPa] & 10 & 10 & 25 \\ 
    $\nu$ & 0.25 & 0.25 & 0.25 \\ 
    $\mu$~[Pa$\cdot$s] & 0.003 & 0.01 & 0.05 \\ 
    $C_l$~[m/s$^1/2$] & 0 & 0 & 0 \\ 
    $Q_0$~[m$^3$/s] & 0.1 & 0.1 & 0.2 \\ 
    $t$~[hrs] & 1 & 1 & 1 \\ 
    $\tau$ & $0.35$ & $15$ & $1.2\times10^4$ \\ 
    $\phi$ & 0 & 0 & 0\\
 \hline
\end{tabular}
\end{center}
\caption{Input parameters for numerical simulations of propagation of multiple hydraulic fractures in the toughness $K$, transition $MK$, and viscosity $M$ regimes.}
\label{tab1}
\end{table}

\begin{figure}[h]
\centering\includegraphics[width=0.85\linewidth]{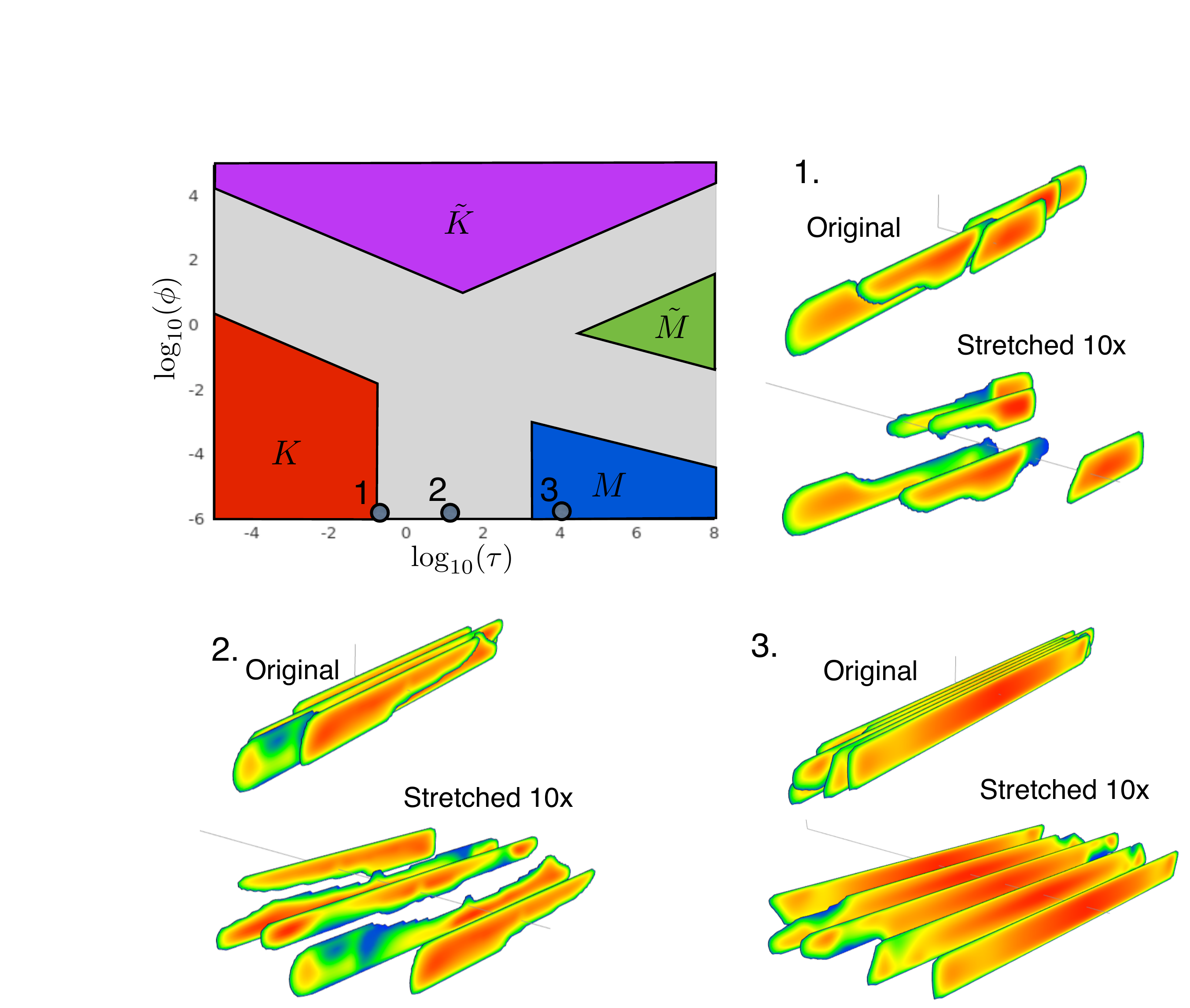}
\caption{Results of numerical simulations for 5 simultaneously propagating hydraulic fractures with 10~m spacing. Parametric space is shown in the top left corner, in which circular markers indicate locations of the problem parameters summarized in Tab.~\ref{tab1}. Actual fracture geometries are shown in two views: original and stretched. In the latter, the result is stretched 10 times along the wellbore to better visualize individual fractures.}
\label{fig3}
\end{figure}

Fig.~\ref{fig3} shows results of numerical simulations for 10~m spacing. First of all, the parametric map (similar to the one shown in Fig.~\ref{figparam}) is shown and three circular markers indicate relative location of the three cases under consideration. As can be seen, one point lies within the toughness domination zone, the second one is in between the toughness and viscosity domination zones, while the third point lies within the viscosity domination zone. Hydraulic fractures for all these three cases are shown in two views: original, in which physical proportions are preserved, and stretched, in which the result is stretched 10 times along the wellbore for better fracture visualization. 

As can be seen from the fracture geometries in Fig.~\ref{fig3}, toughness domination (case 1) leads to unstable fracture growth, whereby the fractures tend to avoid each other and form irregular shapes. Some fractures propagate predominantly to the left from the well and some propagate to the right. Moreover, the ``local'' fracture height of each individual fracture is not equal to the reservoir height (or height of the primary layer). Once the viscosity is increased and toughness is reduced (case 2), the fractures become more symmetric and have nearly identical lengths, even though the ``local'' fracture height is still variable along each individual fracture. Finally, when viscosity dominates (case 3), all the fractures become nearly identical and the fracture height of each fracture is equal to the reservoir height in most regions of the fracture. The exception is the near tip region, whose size is on the order of fracture height $O(H)$, where the non-uniform behavior still persists. Thus, these results demonstrate that the behavior is qualitatively similar to that observed for ``radial'' geometry in~\cite{Dont2020a}, i.e. the case without the barriers restricting vertical growth. One very big difference, however, is the parameter that quantifies the transition from one limiting case to another. For the radial case, the transition from the viscosity to the toughness regime is determined by the parameter
\begin{equation}\label{radialMK}
   4.5\times 10^{-2}< \Bigl(\dfrac{32}{\pi}\Bigr)^{9/2}\dfrac{K_{Ic}^{9}t}{(12^5 \mu^5E'^{13}Q_0^3)^{1/2}}< 2.6\times 10^6,
\end{equation}
where the lower bound corresponds to the onset of the viscosity regime, while the upper bound corresponds to the onset of the toughness regime. For constant height fractures, on the other hand, the result is 
\begin{equation}\label{PKNMK}
0.11<\dfrac{2 \pi^{1/2} E'^4\mu Q_0^2 t}{H^{7/2} K_{Ic}^5}<2.3\times 10^3,
\end{equation}
where the lower bound corresponds to the onset of the toughness regime, while the upper bound corresponds to the onset of the viscosity regime. Clearly, these equations~(\ref{radialMK}) and~(\ref{PKNMK}) are very different and lead to very different results for these geometries.

\begin{figure}[h]
\centering\includegraphics[width=1.0\linewidth]{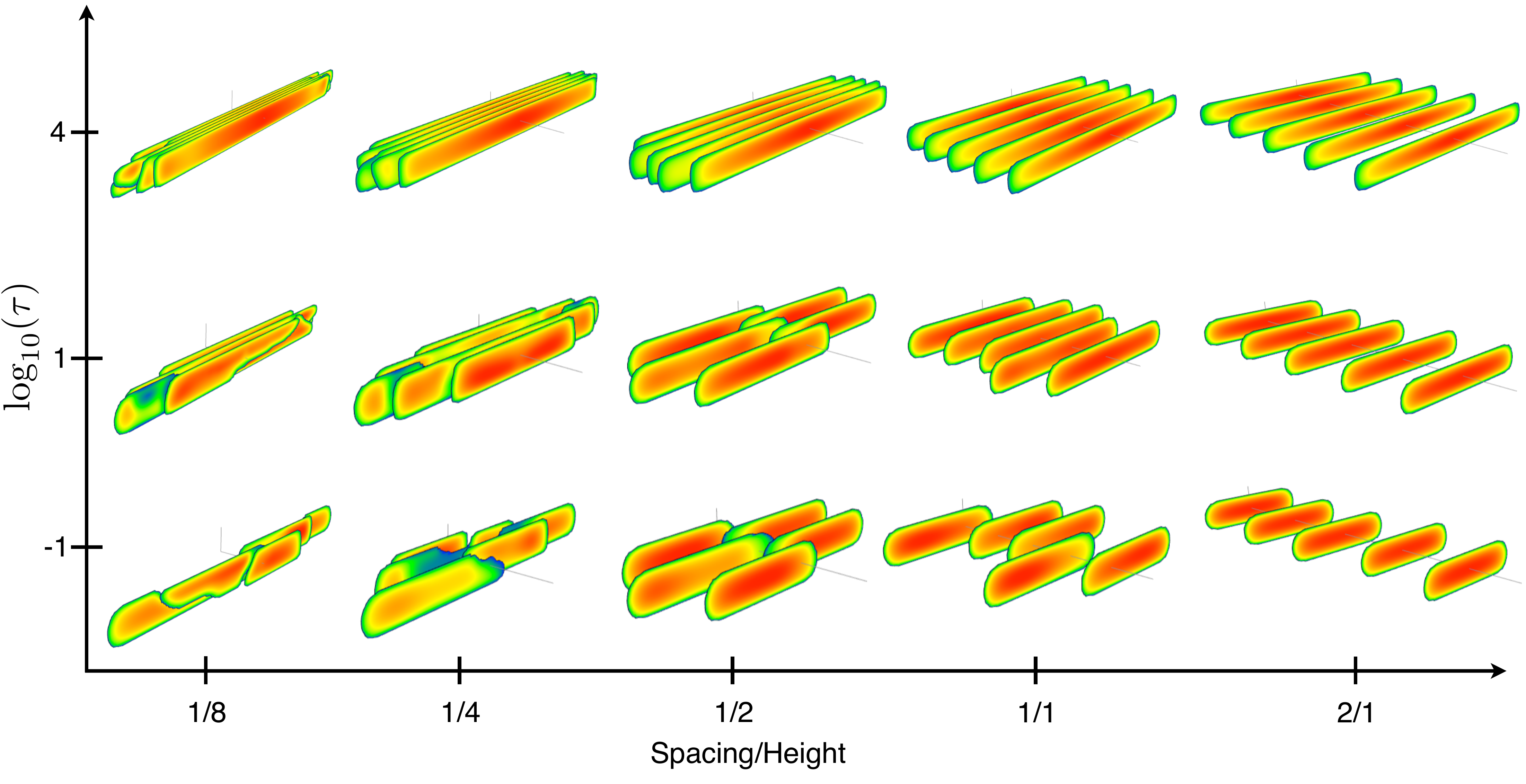}
\caption{Variation of the fracture morphology versus normalized spacing and $\log_{10}(\tau)$ that determines the fracture regime.}
\label{fig4}
\end{figure}

Fig.~\ref{fig4} illustrates sensitivity of the results to fracture spacing. Numerical simulations are performed for the base simulations (see Tab.~\ref{tab1}), but for various spacing ranging from 10~m to 160~m. Given that the reservoir height is 80~m, then the ratios between the spacing and height are from 1/8 to 2/1. The same number of fractures, i.e. 5, is used in all simulations. The vertical axis corresponds to the approximate value of $\log_{10}(\tau)$, but essentially shows the results for toughness (lowest $\tau$), transition (intermediate $\tau$), and viscosity (highest $\tau$) cases. As can be seen from the results, the normalized spacing significantly affects the fracture morphology, as expected. Once the spacing is twice as much as the fracture height, there is almost no interaction between the fractures, which is consistent with the fact that the stress shadow decays quickly at distances commensurate with fracture height (for constant height fractures). Once the fractures are brought closer together, stress interaction starts to distort the fracture shapes if toughness dominates or sufficiently large. Viscosity dominated fractures, on the other hand, remain very similar. These results illustrate how fracture height and spacing, as well as the fracture regime affects the resulting fracture morphology.

\begin{figure}[h]
\centering\includegraphics[width=1.0\linewidth]{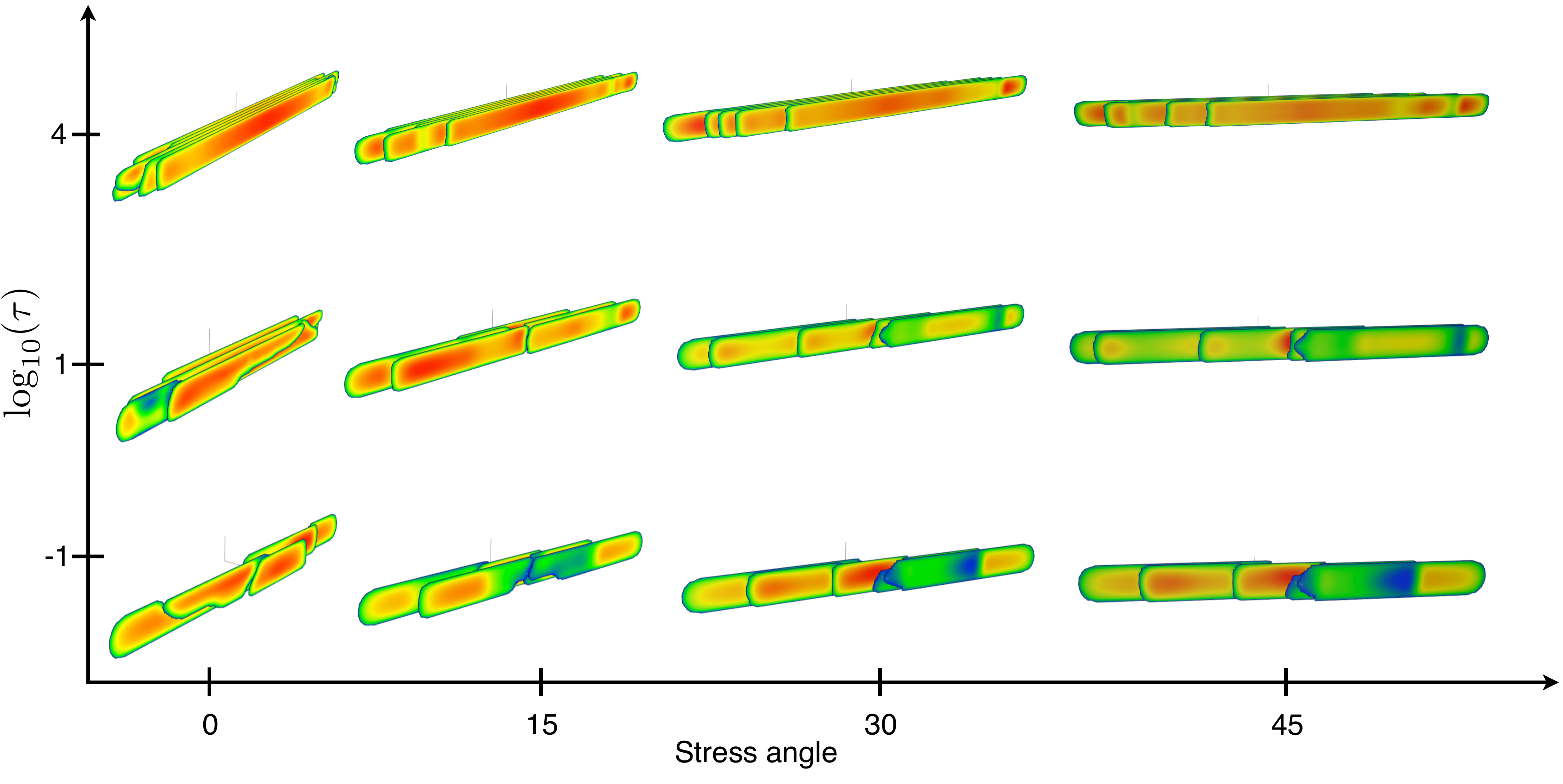}
\caption{Variation of the fracture morphology versus stress orientation angle (in degrees) and $\log_{10}(\tau)$ that determines the fracture regime.}
\label{fig5}
\end{figure}

Finally, Fig.~\ref{fig5} presents sensitivity of the results to stress orientation. In particular, the change of the stress orientation angle allows to effectively adjust the angle between the well and the direction of fracture propagation. The zero angle corresponds to fractures propagating perpendicular to the well, while other cases correspond to the rotation (measured in degrees) relative to this baseline position. Note that the fracture spacing along the well is kept fixed, so that the physical spacing between the fractures, measured perpendicularly to the fracture surface, is effectively multiplied by the cosine of the rotation angle. The input parameters for the ``base'' simulation, i.e. the one corresponding to zero degree stress rotation, are given in Tab.~\ref{tab1}. Results shown in Fig.~\ref{fig5} demonstrate that changing the stress orientation angle noticeably affects the results. First of all, once the symmetry is broken, the fractures that have an initial lateral shift tend to propagate predominantly in the direction of this shift. This applies to all cases, albeit to a different degree. The viscosity dominated cases still have significant fracture overlap, and this overlap drastically reduces towards lower values of $\tau$ or toughness domination. Another interesting observation is that the height of each individual fracture tends to be equal to the reservoir height for larger stress angles, even in the toughness domination regime. That is, the complexity of fracture morphology is reduced once the stress orientation angle increases. This is caused perhaps by the initial bias that fractures have, which significantly affects the stress interaction at early times of propagation. Note that such a behavior persists even if the fluid viscosity is further reduced, i.e. when the system is brought further inside the toughness domination zone.

\section{Summary}\label{secsumm}

In this work, hydraulic fracture morphology for the case of simultaneous propagation of several fractures is investigated in view of dominance of either fluid viscosity or toughness. In particular, the results are obtained for the case of height contained fractures. First, the dimensionless parameters that define the parametric space are outlined and a single parameter that determines the transition from toughness to viscosity domination is defined. Numerical results for 3 base cases (toughness domination, viscosity domination, and transition) are performed for the case of 5 uniformly spaced fractures. It is observed that toughness domination leads to strong stress interaction between the fractures, which leads to the creation of complex fracture geometries. At the same time, relatively large viscosity tends to regularize the fracture morphology and leads to the formation of nearly identical fractures. Numerical simulations with different fracture spacings indicate that the ratio between the fracture spacing to the reservoir height plays a crucial role, since the stress interaction decays quickly as the spacing to height ratio is increased. Also, the influence of stress orientation on the results is investigated, which showed that it also affects the fracture morphology, but to a smaller degree. Since the results are presented with the reference to the dimensionless parametric space, these findings can help to quickly identify the fracture regime and morphology for any input parameters and can be used as a guideline to reach a desired fracture morphology configuration.

\section*{Acknowledgments}
The author would like to acknowledge Mark McClure for useful recommendations regarding this work.



\begin{thebibliography}{10}

\bibitem{King2012}
G.~King.
\newblock Hydraulic fracturing 101: What every representative,
  environmentalist, regulator,reporter, investor, university researcher,
  neighbor and engineer should know about estimatingfrac risk and improving
  frac performance in unconventional gas and oil wells.
\newblock In {\em In Proceedings of SPE Hydraulic Fracturing Technology
  Conference and Exhibition, 6-8 February, The Woodlands, Texas, USA, SPE
  152596}, 2012.

\bibitem{Dont2020a}
E.~Dontsov and R.~Suarez-Rivera.
\newblock Propagation of multiple hydraulic fractures in different regimes.
\newblock {\em Int. J. of Rock Mech. and Min. Sci.}, 128:104270, 2020.

\bibitem{McClure2020}
M.~McClure, M.~Picone, G.~Fowler, D.~Ratcliff, C.~Kang, S.~Medam, and
  J.~Frantz.
\newblock Nuances and frequently asked questions in field-scale hydraulic
  fracture modeling.
\newblock In {\em SPE Hydraulic Fracturing Technology Conference, 4-6 February,
  The Woodlands, Texas, USA, SPE-199726-MS}, 2020.

\bibitem{Detou2016}
E.~Detournay.
\newblock Mechanics of hydraulic fractures.
\newblock {\em Annu. Rev. Fluid Mech.}, 48:31139, 2016.

\bibitem{Perk1961}
T.K. Perkins and L.R. Kern.
\newblock Widths of hydraulic fractures.
\newblock {\em J. Pet. Tech. Trans. AIME}, pages 937--949, 1961.

\bibitem{Nord1972}
R.P. Nordgren.
\newblock Propagation of vertical hydraulic fractures.
\newblock {\em Soc. Petrol. Eng. J.}, pages 306--314, 1972.

\bibitem{Nolte1991}
K.G. Nolte.
\newblock Fracturing-pressure analysis for nonideal behavior.
\newblock {\em J. Pet. Technol.}, 43:210--218, 1991.

\bibitem{Sarva2015}
E.~Sarvaramini and D.~Garagash.
\newblock Breakdown of a pressurized fingerlike crack in a permeable solid.
\newblock {\em J. Appl. Mech}, 82:061006, 2015.

\bibitem{Dont2016}
E.V. Dontsov and A.~P. Peirce.
\newblock Comparison of toughness propagation criteria for blade-like and
  pseudo-3d hydraulic fractures.
\newblock {\em Eng. Frac. Mech.}, 160:238--247, 2016.

\bibitem{Dont2016f}
E.V. Dontsov.
\newblock An approximate solution for a penny-shaped hydraulic fracture that
  accounts for fracture toughness, fluid viscosity, and leak-off.
\newblock {\em R. Soc. open sci.}, 3:160737, 2016.

\bibitem{Dont2022}
E.~Dontsov.
\newblock Analysis of a constant height hydraulic fracture.
\newblock {\em arXiv:2110.13088}, 2022.

\bibitem{McClure2018}
M.~McClure, C.~Kang, C.~Hewson, and S.~Medam.
\newblock Resfrac technical writeup.
\newblock {\em arXiv:1804.02092}, 2018.

\end{thebibliography}


\end{document}